\newcommand{\be}{\begin{equation}}\newcommand{\ee}{\end{equation}}
\newcommand{\bea}{\begin{eqnarray}}\newcommand{\eea}{\end{eqnarray}}
\newcommand{\nn}{\nonumber}\newcommand{\p}[1]{(\ref{#1})}
 \newcommand{\lb}[1]{\label{#1}}
 \newcommand\q{\quad}

\def\a{\alpha}

\def\bt{{\beta}}

\def\g{\gamma}

\def\de{\delta}

\def\eps{\epsilon}

\def\ve{\varepsilon}

\def\la{\lambda}

\def\om{\omega}

 \newcommand\te{\theta}  \def\bth{\bar\theta}

\def\rh{\rho}

\def\z{\zeta}

\def\La{\Lambda}
\def\Om{\Omega}
\def\T{\Theta}

\def\pa{\partial}

\def\na{\nabla}

\newcommand\ab{{\alpha\beta}}

\newcommand\cD{{\cal D}}
\newcommand\cA{{\cal A}}
\newcommand\cN{{\cal N}}
\newcommand\cW{{\cal W}}
\newcommand\cM{{\cal M}}
\newcommand\cR{{\cal R}}
\newcommand\cL{{\cal L}}

\def\sfrac#1#2{{\textstyle\frac{#1}{#2}}}

\documentclass[12pt]{article}
\usepackage{amsfonts}
\def\theequation{\arabic{section}.\arabic{equation}}
\begin{document}
\renewcommand{\thefootnote}{\fnsymbol{footnote}}

\begin{center}
{\large\bf THREE-DIMENSIONAL $\cN=4$ SUPERSYMMETRY  IN HARMONIC
$\cN=3$ SUPERSPACE}
\vspace{0.5cm}

{\bf B.M. Zupnik}
\end{center}

{\it
Bogoliubov Laboratory of Theoretical Physics, JINR, Dubna,
  Moscow Region, 141980, Russia; E-mail: zupnik@theor.jinr.ru}

\begin{abstract}
We consider the map of three-dimensional  $\cN=4$ superfields to
$\cN=3$ harmonic superspace. The left and right representations
of the $\cN=4$ superconformal group are constructed on $\cN=3$
analytic superfields. These representations are convenient for the
description of  $\cN=4$ superconformal couplings of the Abelian gauge
superfields with hypermultiplets. We analyze the $\cN=4$ invariance in
the non-Abelian $\cN=3$ Yang-Mills theory.
\end{abstract}


\setcounter{equation}0

\section{ Introduction}
The simplest supersymmetric Chern-Simons theory was constructed in the
$\cN{=}1, d{=}3 $ superspace with the real coordinates
 $z=(x^m, \te^\a)$, where the Grassmann coordinate  $\te^\a$ has the spinor
index $\a=1, 2$ of the group $SL(2,R)$, and $m=0, 1,2$ is the
three-dimensional vector index \cite{Si,Scho}. This theory uses the
spinor gauge superfield $A_\a(z)$. The superfield action of the $
\cN{=}1$ Chern-Simons theory was interpreted as a superspace
integral of the differential Chern-Simons superform
 $dA+\sfrac{2i}3A^3$ in the framework of the theory of
superfield integral forms \cite{ZP1}-\cite{Z6}. The non-Abelian $\cN{=}2,
d{=}3 $ Chern-Simons action was considered in the superspace
$z=(x^m,\te^\a,\bth^\a)$, where $\te^\a$ and $\bth^\a$ are complex
conjugated spinor coordinates \cite{ZP1,Iv,NG}. The basic
$\cN=2$ gauge superfield is $V(z)$, and the gauge group preserves chirality
of the matter superfields. The $\cN=3, d=3$ Chern-Simons theory was first
studied by the harmonic-superspace method \cite{ZK,Z3}. The analytic gauge
$\cN=3, d=3$ prepotential is similar to the gauge superfield of the
$\cN=2, d=4$ Yang-Mills theory \cite{GIKOS1,GIOS}.

The $\cN=5$ and $\cN=6$ Chern-Simons theories were considered also in
the harmonic-superspace method \cite{HL,Z8}, but this approach
did not succeed in constructing of coupling of the gauge superfield with
matter. We note that the $\cN=6$ Chern-Simons supermultiplet has an
infinite number of auxiliary fields off mass-shell.

The $\cN=6$ Chern-Simons-matter model for the gauge group
$U(N)\times U(N)$ ($ABJM$-model \cite{ABJM}) and the $\cN=8$
$BLG$-model for the gauge group $SU(2)\times SU(2)$ \cite{BLG} were
investigated in the $\cN=3$ harmonic superspace \cite{BILPSZ}. The
$\cN=3$ supersymmetry is manifest in this formalism, the higher
supersymmetry transformations connect different superfields and the
corresponding algebra of transformations closes on the mass shell.
The quantum aspects of the $\cN=3$ superfield theories were analyzed
in \cite{BILPSZ2}.

The $\cN=4, d=3$ superfield theories were studied in our papers
\cite{Z5,Z6,Z7}.
The mirror left and right $\cN=4$ supermultiplets are defined in
different harmonic superspaces and this fact is the main obstacle
to the construction of the left-right couplings.
We review the $\cN=4, d=3$ superspace formulas in Appendix.

In the next section, we analyze the $\cN=3, d=3$ superconformal
transformations in the standard and harmonic superspaces.  The
corresponding Killing operator
$K_3$ contains the generators of the $\cN=3$ superconformal group
$P_m, M_m, D, V_{kl}, Q^{(kl)}_\a$
and $S^{(kl)}_\a$ acting as  superspace differential operators.

We study the relations between the $\cN=4$ and $\cN=3$ harmonic superfields
in section 3. The left $\cN=4$ analytic superfields are decomposed in terms
of the $\cN=3$ spinor coordinates $\te^{++\a}, \te^{0\a}$ and the
additional spinor coordinate $\te^\a_4$. We obtain the $\cN=3$
representation of the left $\cN=4$ supermultiplets using the operator
$O$ connecting different superspaces.
This operator allows us to define the active
$\cN=4$ superconformal transformations on the analytic $\cN=3$ superfields.
These transformations include the $\cN=3$ superconformal generators
and the transformations with the additional even generator $A_{kl}$
and the odd spinor generators $Q^4_\a$ and $S^4_\a$. The mirror map $\cM$
from the left to right $\cN=4$ representations is equivalent to the
change  of the signs of  these additional generators. Thus, we study
the left and right supermultiplets in the same analytic superspace,
the mirror supermultiplets have similar $\cN=3$ transformations, and their
additional transformations differ by the signs.

Section 4 is devoted to the construction of the $\cN=4$ superconformal
models in the $\cN=3$ harmonic superspace. We consider the  analytic
Abelian gauge prepotential $V^{++}_0$ and its analytic superfield
strength $W^{++}_0$ and prove that these superfields have the mirror
$\cN=4$ transformations. The $\cN=4$ superconformal $BF$ coupling of
$W^{++}_0$ contains the right Abelian prepotential $A^{++}$. This
coupling is equivalent to the difference of two Abelian Chern-Simons
interactions. The natural $\cN=4$ superconformal coupling of the
left gauge superfield $V^{++}_0$ is defined with the left
hypermultiplets, while the right $\cN=4$ gauge multiplet $A^{++}$
interacts with the right hypermultiplets. We consider also the
nonminimal interaction of $W^{++}_0$ with the right hypermultiplet.

The interesting $\cN=4$ superconformal coupling $S^\bt(V^{++}_0,A^{++})$
of two Abelian superfields contains the $BF$ coupling and the improved
forms of the actions for two superfield strengths. Analogous improved
forms were considered in the $\cN=2, d=4$ superspace \cite{GIOS} and in
the $\cN=4, d=3$ superspaces \cite{Z7}. The $\cN=3$ representation
is convenient for the analysis of the quantum properties of this model.
We also discuss the action of the nonlinear $\cN=4$ electrodynamics.

In section 5 we study the $\cN=4$ superconformal transformations of the
$\cN=3$ gauge superfields. In particular, we obtain the nonlinear fourth
supersymmetry transformation of the non-Abelian $\cN=3$ superfield
strength and prove the $\cN=4$ invariance of the corresponding non-Abelian
action.

\setcounter{equation}0
\section{ $\cN=3, d=3$ harmonic superspace}
We use the notation of papers \cite{Z3,BILPSZ} in the $\cN=3$
superspace. The real coordinates of the  $\cN=3$ superspace are
introduced in the central basis \bea &&z=(x^m, \te^\a_{(kl)}),\q
\overline{\te^\a_{(kl)}}=\te^{(kl)\a},\lb{CB} \eea where $k,l$ are
the spinor indices of the automorphism group $SU_V(2)$. The
isovector representation for the spinor coordinates contains the
Pauli matrices $\te_B^\a=\frac{i}{\sqrt2}(\tau_B)^{kl}
\te^\a_{(kl)},\q B=1, 2, 3$. The spinor derivatives in these
coordinates are \bea
&&D^{(kl)}_\a=\pa^{(kl)}_\a+i\te^{(kl)\bt}\pa_\ab,\q
\pa_\ab=(\g^m)_\ab\pa_m,
\nn\\
&&\pa_m x^n=\de^n_m,\q \pa^{(kl)}_\a\te^\bt_{(jn)}=\frac12(\de^k_j\de^l_n+
\de^l_j\de^k_n),
\eea
where $(\g^m)_\ab$ are the three-dimensional  $\gamma$ matrices.
The transformations of the $\cN=3$ superconformal group in the
real coordinates \p{CB} \cite{BILPSZ} can be rewritten via the Killing
superconformal operator $K_3$ that contains the generators of the
corresponding Lie superalgebra
\bea
&&\de_{sc} x^m=K_3x^m,\q
\de_{sc} \te^\a_{(kl)}=K_3\te^\a_{(kl)},\nn\\
&&K_3(z)=c^mP_m+l^mM_m+bD+k^mK_m+a^{kl}V_{kl}
+\eps^\a_{(kl)}Q^{(kl)}_\a+\eta^\a_{(kl)}S^{(kl)}_\a,\lb{CKil}
\eea
where $c^m, l^m, b, k^m, a^{kl}, \eps^\a_{(kl)}$ and $\eta^\a_{(kl)}$ are
the superconformal parameters. It is easy to construct these superconformal
generators in the central basis, for instance, the $SU_V(2)$ generators have
the form
\bea
V_{kl}=\ve_{ln}\te^\a_{(kj)}\pa^{(jn)}_\a+\ve_{kn}\te^\a_{(lj)}\pa^{(jn)}_\a.
\lb{3V}
\eea

The Killing operator connects the active superconformal transformations
of superfields $\de^*\cA$ with the passive superconformal transformations
\bea
\de_{sc}\cA(z)\equiv \cA(z+\de_{sc}z)-\cA(z)=\de^*\cA(z)+K_3\cA(z).
\eea

We consider the important commutation relations of the Killing operator with
the spinor derivatives, which define the passive superconformal
transformations of these derivatives $\de_{sc}D^{(kl)}_\a$
\bea
&&[K_3,D^{(kl)}_\a]=-\frac12jD^{(kl)}_\a-\chi_\a^\g D^{(kl)}_\g+\la^k_n
D^{(nl)}_\a+\la^l_n D^{(kn)}_\a=\de_{sc}D^{(kl)}_\a\lb{K3D}
\eea
where
\bea
&&\chi^{\a\g}=a^{\a\g}+\frac14(x^{\a\rh}k_\rh^\g+x^{\g\rh}k_\rh^\a)
-\frac{i}4\te^\bt_{(kl)}\te^{(kl)}_\bt
k^{\a\g}+\frac{i}2(\te^\a_{(kl)}
\eta^{(kl)\g}+\te^\g_{(kl)}\eta^{(kl)\a}),\nn\\
&&\la^{kl}=a^{kl}-\frac{i}2\ve_{jn}\te^{kj\a}\te^{nl\bt}k_\ab-
\frac{i}2\ve_{jn}(\te^{kj\a}\eta^{nl}_\a+\te^{lj\a}\eta^{nk}_\a),
\lb{CBla}\\
&&j=-b-k_mx^m-i\te^\a_{(kl)}\eta^{(kl)}_\a.\lb{CBpar}
\eea
are the superfield parameters. We consider the useful relations
\bea
&&D^{(kl)}_\g\chi^\rh_\a=-\de^\rh_\g D^{(kl)}_\a j
+\frac12\de^\rh_\a D^{(kl)}_\g j,\nn\\
&&D^{(ni)}_\a \la^{kl}=\frac14\ve^{nk}D^{(il)}_\a j+\frac14\ve^{ik}
D^{(nl)}_\a j
+\frac14\ve^{il}D^{(kn)}_\a j+\frac14\ve^{nl}D^{(ik)}_\a j.\lb{SC10}
\eea

The primary superfield of the weight $w$ has the active superconformal
transformation
\bea
&&\de^*A_w=wj A_w-K_3A_w.
\eea
We study transformations of the spinor derivative of superfield  $A_w$
using the formal relation
$[\de^*,D^{(kl)}_\a]=0$ and  formula \p{K3D}.

The $\cN=3$ superconformal linear (tensor) multiplet $W^{(kl)}$ satisfies
the following superfield constraints:
\bea
D^{((jn)}_\a W^{(kl))}=0,\lb{LM}
\eea
where the brackets mean symmetrization in four indices. These constraints
are covariant under
the superconformal transformations
\bea
\de^*W^{(kl)}=jW^{(kl)}+\la^k_nW^{(nl)}+\la^l_nW^{(kn)}-K_3W^{(kl)}.\lb{CBW}
\eea

The superconformal transformation of the $SU_V(2)/U(1)$ harmonics
$u^\pm_i$  can be defined as follows:
\bea
&&\de_{sc}u^+_i=\la^{kl}u^+_ku^+_lu^-_i,\q \de_{sc}u^-_i=0
\eea
where the matrix $\la^{kl}$ is given by \p{CBla}.
The harmonic projection of the
relation \p{SC10} yields the condition
\bea
&&u^+_nu^+_iu^+_ku^+_lD^{(ni)}_\a \la^{kl}=D^{++}_\a\la^{++}=0,\q \la^{++}=
\la^{kl}u^+_ku^+_l,\q D^{++}_\a=u^+_nu^+_iD^{(ni)}_\a.
\eea
The Killing vector $K^C_3=K_3(z)+\la^{++}\pa^{--}$ in the extended central
basis $(z, u)$ preserves the Grassmann analyticity condition
$D^{++}_\a A(z,u)=0$
\bea
&&[K^C_3,D^{++}_\a]=-\chi_\a^\g D^{++}_\g-\frac12(j-\pa^{--}\la^{++})
D^{++}_\a.
\eea

The  Grassmann analyticity is manifest in
the analytic basis $z_A=(\z, \te^{--\a})$
\bea
&&\z=(x^m_A, \te^{++\a}, \te^{0\a}, u^\pm_i),\\
&&x^m_A=x^m+i(\g^m)_\ab \te^{++\a}\te^{--\bt},\q
\te^{\pm\pm\a}=U^{\pm\pm kl} \te^\a_{(kl)},\q \te^{0\a}=U^{0kl}
\te^\a_{(kl)},\lb{3AB}
\eea
where we define the isovector combinations of
the spinor harmonics
\bea
&&U^{++kl}=u^{+k}u^{+l},\q U^{--kl}=u^{-k}u^{-l},\q
U^{0kl}=\frac12(u^{+k}u^{-l} +u^{+l}u^{-k}),\\
&&\int du U^{++kl}U^{--}_{ij}=-2\int du U^{0kl}U^{0}_{ij}=\frac16(
\delta^k_i\delta^l_j+\delta^k_j\delta^l_i).
\eea
The special conjugation $\sim$ is defined in the analytic coordinates
\bea
&&\widetilde{u^{\pm}_i}=u^{\pm i},\q \widetilde{x^m_A}=x^m_A,\q
 \widetilde{\te^{0,\pm\pm}_\a}=\te^{0,\pm\pm}_\a.
\eea

In the analytic basis, the harmonic and spinor derivatives have the form
\bea
&&\cD^{++}=\pa^{++}+2i\te^{++\a}\te^{0\bt}\pa^A_\ab+\te^{++\a}\pa^0_\a
+2\te^{0\a}\pa^{++}_\a,\q\pa^A_\ab=(\g^m)_\ab
\pa^A_m,\nn\\
&&\cD^{--}=\pa^{--}-2i\te^{--\a}\te^{0\bt}\pa^A_\ab+\te^{--\a}\pa^0_\a
+2\te^{0\a}\pa^{--}_\a,\nn\\
&&\cD^0=\pa^0+2\te^{++\a}\pa^{--}_\a-2\te^{--\a}\pa^{++}_\a,
\q [\cD^{++},\cD^{--}]=\cD^0,\\
&&D^{++}_\a=\pa^{++}_\a,\q D^{--}_\a=\pa^{--}_\a+2i\te^{--\bt}\pa^A_\ab,\q
D^0_\a=-\frac12\pa^0_\a+i\te^{0\bt}\pa^A_\ab,\\
&&\pa_m^Ax^n_A=\de^n_m,\q\pa^0_\a\te^{0\bt}=\de^\bt_\a,\q \pa^{\mp\mp}_\a
\te^{\pm\pm\bt}=\de^\bt_\a,\q [\pa^{++},\pa^{--}]=\pa^0,\nn
\eea
where $\pa^0, \pa^{\pm\pm}$ are the partial harmonic derivatives.
We use the notation
\bea
&(\te^{++})^2=\te^{++\a}\te^{++}_\a,\q (\te^{0})^2=\te^{0\a}\te^{0}_\a,\q
(\te^{++}\te^0)=\te^{++\a}\te^{0}_\a,\q (\te^{++}\g^m\te^0)=
\te^{++\a}\te^{0\bt}(\g^m)_\ab,&\nn\\
&(D^{++})^2=D^{++\a}D^{++}_\a,\q (D^{0})^2=D^{0\a}D^{0}_\a.&
\eea

The $\cN=3$ analytic superfield $\phi(\z)$ satisfies the condition
$D^{++}_\a\phi=0$.
The $\cN=3$ superconformal transformations of the analytic coordinates
\cite{BILPSZ} are generated by  the Killing operator
\bea
&&K^A_3={\cal K}_3(\z)+\de_{sc}\te^{--\a}\pa^{++}_\a,\lb{K3A}
\eea
where the operator ${\cal K}_3(\z)$ acts in the analytic subspace
$\z$, for instance,
\bea
&&{\cal K}_3u^+_i=\la^{++}(\z,u)u^-_i,\q {\cal K}_3u^-_i=0,\\
&&\la^{++}(\z)=-i(\te^{++}\g_m\te^0)k^m+i(\te^{0\a}U^{++}_{kl}
-\te^{++\a}U^{0}_{kl})\eta^{(kl)}_\a+U^{++}_{kl}a^{kl}.\lb{lapp3}
\eea

The  analytic representation of the even superconformal $\cN=3$
generators has the form
\bea
&&P_m=\pa_m^A,\q M_m=\ve_{mnp}x^n_A\pa^{Ap}
+\frac12(\g_m)_\a^\bt(\te^{++\a}\pa^{--}_\bt+\te^{0\a}\pa^0_\bt),\\
&& D=x^m_A\pa^A_m+\frac12(\te^{++\a}\pa^{--}_\a+\te^{0\a}\pa^{0}_\a),
\\
&& K_m=x_{Am}x^p_A\pa^A_p-\frac12(x_A)^2\pa^A_m-i(\te^{++}\g_m\te^0)
\pa^{--}\nn\\
&&-\frac{i}2(\g_m)_\bt^\a(\te^0)^2\te^{++\bt}\pa^{--}_\a
+\frac12x^n_A(\g_n)^\ab(\g_m)_{\bt\g}(\te^{++\g}\pa^{--}_\a
+\te^{0\g}\pa^0_\a),\lb{3KA}\\
&&V_{kl}=U^{++}_{kl}\pa^{--}+U^{--}_{kl}
\te^{++\a}(\pa^0_\a+2i\te^{0\g}\pa^A_{\a\g})
+2U^0_{kl}\te^{++\a}\pa^{--}_\a.\lb{V3}
\eea
The odd analytic superconformal generators act also in the analytic
superspace
\bea
&&Q^{(kl)}_\a=U^{++kl}\pa^{--}_\a-2iU^{--kl}\te^{++\g}\pa^A_{\a\g}
+U^{0kl}(\pa^0_\a+2i\te^{0\g}\pa^A_{\a\g}),\lb{QA}\\
&&S^{(kl)}_\a=\frac13(\g^m)^\bt_\a[K_m,Q^{(kl)}_\bt].
\eea

We consider the active realization of the $\cN=3$ superconformal
transformations on the analytic superfields
\bea
&&\de^*_3\phi=\de_{sc}\phi-{\cal K}_3(\z)\phi.\lb{K3A}
\eea

The Killing operator satisfies the relations
\bea
&&[K^A_3,\cD^{++}]=-\la^{++}\cD^0=\de_{sc}\cD^{++},\nn\\
&& [K^A_3,\cD^{--}]=-(\cD^{--}\la^{++})
\cD^{--}=\de_{sc}\cD^{--},\lb{KD}
\eea
which determine the passive operator transformations.

The harmonic projection of the $\cN=3$ linear multiplet \p{LM}
$W^{++}=u^+_ku^+_lW^{(kl)}$
satisfies the constraints
\bea
&&D^{++}_\a W^{++}=0,\q \cD^{++} W^{++}=0
\eea
and has the superconformal transformation
\bea
&&\de^*_3W^{++}=2\la W^{++}-{\cal K}_3W^{++},\nn\\
&&\la=-\frac12b-\frac12k_mx^m_A+i(\te^{0\a}U^0_{kl}
-\te^{++\a}U^{--}_{kl})\eta^{(kl)}_\a+U^0_{kl}a^{(kl)}.\lb{la3}
\eea

The superfield superconformal parameters $j$ and
$\la^{(kl)}$ \p{CBW} can be expressed via the analytic superconformal
parameters
\bea
&&\la^{kl}=U^{--kl}\la^{++}-U^{0kl}\cD^{--}\la^{++}+\frac12U^{++kl}
(\cD^{--})^2\la^{++},\nn\\
&&j=2\la-\cD^{--}\la^{++},\q \la=-\frac12b+U^0_{kl}\la^{kl}.
\eea

The Abelian scalar analytic  prepotential $V^{++}_0(\z)$ has the
gauge transformation
\bea
&&\de_\La V^{++}_0=-\cD^{++}\La,\q \widetilde{V^{++}_0}=-V^{++}_0,\q
\widetilde{\La}=-\La.\lb{V3}
\eea

The gauge-invariant real  analytic superfield strength can be defined via
this gauge superfield
\cite{Z3,BILPSZ}
\bea
&&W^{++}_0(\z)=-\frac14(D^{++})^2V^{--}_0=IV^{++}_0=\int d\z^{-4}_2
I(\z,\z_2)V^{++}_0(\z_2),\nn\\
&&I(\z,\z_2)=\frac1{16}
(D^{++})^2(D^{++}_2)^2\de^9(z-z_2)\frac{1}{(u^+u^+_2)^2},\lb{W3}
\eea
where $V^{--}_0$ is the harmonic connection, and $I$ is the integral
operator. We use the relations
\be
\cD^{++}V^{--}_0=\cD^{++}V^{++}_0,\q \cD^{++}W^{++}_0=0.
\ee

We note that the superfields $V^{++}_0$ and $W^{++}_0$ have opposite
$P$-parities with respect to the transformation
\bea
P(x^0_A, x^1_A, x^2_A)=(x^0_A, -x^1_A, x^2_A),\q
P\te^{0,\pm\pm}_\a=-(\g_1)_\a^\bt\te^{0,\pm\pm}_\bt.
\eea

The component fields of the vector multiplet $\phi^{kl}, A_m, \la^\a,
\chi^\a_{kl}$ and $X^{kl}$
can be determined in the $WZ$-gauge
\bea
&&V^{++}_{WZ}=3(\te^{++})^2U^{--}_{kl}\phi^{kl}+2(\te^{++}\g^m\te^0)A_m
+2i(\te^0)^2\te^{++\a}\la_\a+3i(\te^{++})^2\te^{0\a}U^{--}_{kl}
\chi_\a^{kl}\nn\\
&&+3i(\te^{++})^2(\te^0)^2U^{--}_{kl}X^{kl}.\lb{VWZ}
\eea
The $P$-even superfield $V^{++}_0$ contains
the pseudoscalar field $\phi^{kl}$, the scalar $X^{kl}$ and the vector field
$A_m$.

The $\cN=3$ supersymmetry transformations of the component fields
\bea
&&\de_\eps\phi^{kl}=-i\eps^{(kl)\a}\la_\a-\frac{i}2(\eps^{(kj)\a}\chi^l_{j\a}
+\eps^{(lj)\a}\chi^k_{j\a}),\\
&&\de_\eps A_m=\frac{i}2(\eps_{(kl)}\g_m\chi^{kl}),\q
\de_\eps\la_\a=-\eps^\g_{(kl)}\pa_{\a\g}\phi^{(kl)}-\eps^{(kl)}_\a X_{kl},\\
&&\de_3\chi^{kl}_\a=
\eps^{(kj)\g}\pa_{\a\g}\phi^{l}_j+\eps^{(lj)\g}
\pa_{\a\g}\phi^{k}_j+(\g_m)_\ab F^m\eps^{\bt(kl)}
-\eps^{(kj)}_\a X^l_j-\eps^{(lj)}_\a X^k_j,\\
&&F^m=\ve^{mnp}(\pa_nA_p-\pa_pA_n),\nn\\
&&\de_\eps X^{kl}=-i\eps^{(kl)\a}\pa_{\a\g}\la^\g
+i(\eps^{(kj)\a}\pa_{\a\g}\chi^{l\g}_j+\eps^{(lj)\a}\pa_{\a\g}
\chi^{k\g}_j)
\eea
can be obtained from the superfield transformation of $V^{++}_{WZ}$.

The Abelian superfield strength has the simple form in this gauge
\bea
&&W^{++}_0=[-(D^0)^2+\frac12(D^0)^2\cD^{--}\cD^{++}+\frac12(D^0D^{--})
\cD^{++}-\frac1{12}(D^0)^2(\cD^{--})^2(\cD^{++})^2\nn\\
&&-\frac16(D^0D^{--})\cD^{--}(\cD^{++})^2-\frac1{24}(D^{--})^2(\cD^{++})^2]
V^{++}_{WZ}.\lb{Wabel}
\eea

The component decomposition of the superfield strength contains the fields
and their derivatives
\bea
&&W^{++}_0=U^{++}_{kl}\phi^{kl}-2i(\te^{++}\g^m\te^0)U^0_{kl}\pa^A_m\phi^{kl}
+(\te^{++})^2(\te^0)^2U^{--}_{kl}\Box\phi^{kl}\nn\\
&&-i(\te^{++}\g_m\te^0)F^m+i\te^{++\a}\la_\a
-(\te^{0})^2\te^{++\a}\pa^A_{\a\g}\la^\g\nn\\
&&+i(\te^{++\a}U^0_{kl}-i\te^{0\a}U^{++}_{kl})\chi^{kl}_\a
+[(\te^{0})^2\te^{++\a}U^{0}_{kl}+(\te^{++})^2\te^{0\a}U^{--}_{kl}]
\pa^A_{\a\g}\chi^{kl\g}\nn\\
&&+i[(\te^{++})^2U^{--}_{kl}+(\te^{0})^2U^{++}_{kl}
-2(\te^{++}\te^0)U^0_{kl}]X^{kl},
\eea
where $\Box=\eta^{mn}\pa^A_m\pa^A_n$ and $F^m$ is the Abelian field strength.

\setcounter{equation}0
\section{Relations between  $\cN=4$ and $\cN=3$ harmonic
superfields }
\subsection{Left  $\cN=4$ representations}

The extended $\cN=3$ analytic basis includes coordinates \p{3AB}
and the additional spinor coordinate $\te^\a_4$
\bea
&&x^m_A, \te^{\pm\pm\a}, \te^{0\a}, \te^\a_4, u^\pm_k.\lb{N34}
\eea
We analyze the relation between the left analytic $\cN=4$ basis
\p{LAB} and the extended  $\cN=3$ basis
\bea
&&\te^{+-\a}=\te^{0\a}+\frac12\te^\a_4,\q \te^{-+\a}=\te^{0\a}-\frac12\te^\a_4,
\lb{altsp}\\
&&x^m_L=x^m_A+i(\g^m)_\ab\te^{0\a}\te^\bt_4.\lb{altx}
\eea

We consider the relations between the differential operators
of two bases
\bea
&&\frac{\pa}{\pa x^m_L}=\frac{\pa}{\pa x^m_A},\q
\pa^{+-}_\a=\frac12\pa^0_\a-
i\te^{0\bt}\pa^A_\ab-\pa^4_\a-\frac{i}2\te^\bt_4\pa^A_\ab,\\
&&\pa^{-+}_\a=\frac12\pa^0_\a+i\te^{0\bt}\pa^A_\ab+\pa^4_\a
-\frac{i}2\te^\bt_4\pa^A_\ab,\\
&& D^{++}_L=\cD^{++},\q D^{--}_L=\cD^{--}.
\eea
The partial harmonic derivatives of different bases are identical, because
the coordinate transformations \p{altsp}-\p{altx} do not contain harmonics
manifestly. The $\cN=4$ supersymmetry generators \p{QA4} can be
rewritten in the basis \p{N34}
\bea
&&Q^4_\a=\pa^4_\a-\frac{i}2\te^\bt_4\pa^A_\ab,
\eea
and operators $Q^{(kl)}_\a$ are given by the standard formula \p{QA}.

The left analytic $\cN=4$ superfield $\hat\Phi_L$ in the extended
$\cN=3$ analytic basis
\bea
&&\Phi_L(x^m_L,\te^{++\a},\te^{+-\a},u)=O\phi_L(x^m_A,\te^{++\a},
\te^{0\a},u),\lb{L43}\\
&&O=\exp(-\te^\a_4D^0_\a)=1-\te^\a_4D^0_\a-\frac14(\te_4)^2(D^0)^2
\eea
is connected by the invertible operator $O$ to the corresponding $\cN=3$
analytic superfield $\phi_L(\z)$. It is evident, that the
$\cN=4$ and $\cN=3$ analytic superspaces have equal dimensions, and their
differential operators are connected by the
 $O$-map
\bea
&&O^{-1}\pa^{+-}_\a O=-\pa^4_\a,\q O^{-1}\pa^{-+}_\a O=\pa^0_\a+\pa^4_\a,
\q O^{-1}\pa_m^LO=\pa_m^A,\q O^{-1}\pa^{\pm\pm} O=\pa^{\pm\pm},\nn\\
&&O^{-1}x^m_LO=x^m_A,\q O^{-1}\te^{+-\a}O=\te^{0\a},\q
O^{-1}\te^{-+\a}O=\te^{0\a}-\te^\a_4.\lb{Otr}
\eea
The representation \p{L43} has the manifest left analyticity.

The $O$-transformations of the  $\cN=4$ supersymmetry
generators and the harmonic derivatives give us
\bea
&&O^{-1}Q^{(kl)}_\a O=Q^{(kl)}_\a,\q O^{-1}Q^4_\a O={\cal Q}^4_\a+\pa^4_\a,
\q{\cal Q}^4_\a=-D^0_\a,\\
&&O^{-1}D^{\pm\pm}_L O=\cD^{\pm\pm}_L=\cD^{\pm\pm}-\te^\a_4D^{\pm\pm}_\a,\q
[\cD^{\pm\pm}_L,({\cal Q}^4_\a+\pa^4_\a)]=0.
\eea

The $O$-transformations of the $SU_L(2)\times SU_R(2)$
generators \p{4LL},\p{4LR} can be obtained analogously
\bea
&&O^{-1}L_{kl}O\phi_L=\cL_{kl}\phi_L,\q
O^{-1}R_{kl}O\phi_L=\cR_{kl}\phi_L,\lb{OLR}
\eea
where the analytic parts of the operators have the form
\bea
&&\cL_{kl}=U^{++}_{kl}\pa^{--}
+U^{++}_{kl}\te^{0\a}\pa^{--}_\a
+U^0_{kl}(\te^{++\a}\pa^{--}_\a+\te^{0\a}\pa^{0}_\a)
+2iU^{--}_{kl}\te^{++\a}\te^{0\a}\pa^A_\ab,\lb{Lan}\\
&&\cR_{kl}=-U^{++}_{kl}\te^{0\a}\pa^{--}_\a+U^0_{kl}\te^{++\a}\pa^{--}_\a
-U^0_{kl}\te^{0\a}\pa^{0}_\a
+U^{--}_{kl}\te^{++\a}\pa^{0}_\a.\lb{Ran}
\eea

The map  \p{L43} defines the representation of the
$\cN=4$ superconformal group on the $\cN=3$ superfield $\phi_L(\z)$
\bea
&&K_4\Phi_L(\hat\z_L)~\rightarrow~{\cal K}_L\phi_L(\z)=O^{-1}K_4\,O\,
\phi_L,\\
&&{\cal K}_L={\cal K}_3(\z)+{\cal K}_{4/3}(\z),\q {\cal K}_{4/3}=
\eps^\a_4{\cal Q}^4_\a
+\eta^\a_4{\cal S}^4_\a+b^{kl}\cA_{kl}\lb{add}
\eea
where ${\cal K}_3(\z)$ is the analytic representation of the
$\cN=3$ Killing operator \p{K3A}, and the additional operators in
 ${\cal K}_{4/3}$ have the form
\bea
&& {\cal Q}^4_\a=-\,D^0_\a,\q {\cal S}^4_\a=\frac13(\g^m)_\ab
[K_m,{\cal Q}^{4\bt}],\\
&&{\cal A}_{kl}=\cL_{kl}-\cR_{kl}=U^{++}_{kl}(\pa^{--}+2\te^{0\a}\pa^{--}_\a)
+2U^0_{kl}\te^{0\a}\pa^0_\a
\nn\\
&&
-U^{--}_{kl}\te^{++\a}(\pa^{0}_\a+2i\te^{0\bt}
\pa^A_\ab)\lb{cA}.
\eea

It is easy to check the relation
\bea
&&[\cA_{kl},{\cal Q}_\a^4]=-[\cA_{kl},D^0_\a]=-Q_{(kl)\a}.
\eea

We use the commutation relation
\bea
&&[{\cal K}_{4/3},\cD^{++}]=-\la^{++}_4\cD^0,\nn\\
&&\la^{++}_4=-\frac{i}2\eta^\a_4\te^{++}_\a+b^{kl}U^{++}_{kl}
=\cD^{++}\la_4,\q
\la_4=-\frac{i}2\eta^\a_4\te^{0}_\a+b^{kl}U^0_{kl}.\lb{lapp4}
\eea
In the $\cN=3$ representation, the $\cN=4$ superconformal transformation of
the left tensor multiplet $L^{++}$ has the form
\bea
&&\de^*_4L^{++}=2(\la+\la_4)L^{++}-{\cal K}_LL^{++}.\lb{ltens}
\eea

\subsection{ Right   $\cN=4$ representations}

The alternative basis in the right $\cN=4$ superspace \p{RAB}
contains $x^m_R$ and the spinor coordinates
 $\te^{\pm\pm}_\a=u^\pm_ku^\pm_l\te^{(kl)}_\a,\q
\te^{\pm\mp}_\a=u^\pm_ku^\mp_l\te^{(kl)}_\a$. We consider the
relation between the right $\cN=4$ superspace and the extended
 $\cN=3$ basis \p{N34}
\bea
&&x^m_R=\cM x^m_L=x^m_A-i(\g^m)_\ab\te^{0\a}\te^\bt_4.
\eea

The action of the map $\cM$ on the left $\cN=4$ analytic superfield
in the extended $\cN=3$ basis changes the sign of  $\te^\a_4$ in the
representation \p{L43} and gives the formula for the right analytic
superfield
\bea
&&\cM\Phi_L=\Phi_R(x^m_R,\te^{++\a},\te^{-+\a},u)
=O^{-1}\phi_R(x^m_A,\te^{++\a},
\te^{0\a},u),\lb{R43}\\
&&O^{-1}=\cM O=\exp(\te^\a_4D^0_\a)=1+\te^\a_4D^0_\a
-\frac14(\te_4)^2(D^0)^2,
\eea
where $\phi_R(\z)$ is the corresponding $\cN=3$ analytic superfield.

The representation \p{R43} guarantees the manifest right analyticity
\bea
&&\pa^{-+}_\a[O^{-1}\phi_R(\z)]=0.
\eea

The mirror map of  formulae \p{OLR} yields the right representation of
the $SU_L(2)\times SU_R(2)$ generators
\bea
&&O\hat{L}_{kl}O^{-1}\phi_R=\cM \cL_{kl}\phi_L=\cR_{kl}\phi_R,\nn\\
&&O\hat{R}_{kl}O^{-1}\phi_R=\cM \cR_{kl}\phi_L=\cL_{kl}\phi_R,
\eea
where the analytic operators are given in \p{Lan} and \p{Ran}.

The mirror map of the   $\cN=4$ superconformal group has the form
\bea
&&K_4\Phi_R(\hat\z_R)~\rightarrow~{\cal K}_R\phi_R(\z),\\
&&{\cal K}_R=\cM {\cal K}_L={\cal K}_3(\z)-{\cal K}_{4/3},
\eea
where ${\cal K}_{4/3}(\z)$ includes the additional $\cN=4$ generators \p{add}.
Thus, the map $\phi_R(\z)=\cM \phi_L$ changes the signs in transformations
with the additional $\cN=4$ generators, for instance, $\cM \eps^\a_4
D^0_\a\phi_L=-\eps^\a_4D^0_\a\phi_R$.

\setcounter{equation}0
\section{ $\cN=4$ superconformal models in $\cN=3$\\
 superspace}
 The manifestly supersymmetric $\cN=4$ models were investigated
 in the $\cN=4$ superspace \cite{Z7}. We reformulate these models
 in the  $\cN=3$ basis (\ref{N34}). The corresponding representation
 of the left Abelian
${\cal N}=4$ gauge superfield is
\begin{eqnarray}
&&V^{++}_{4L}=[1-\te^\a_4D^0_\a-\frac14(\te_4)^2(D^0)^2]V^{++}_0(\z),\label{V4}
\end{eqnarray}
where $V^{++}_0(\z)$ is the Abelian ${\cal N}=3$ analytic prepotential
(\ref{V3}). The nonanalytic ${\cal N}=4$ Abelian harmonic connection
satisfies the equation
\begin{eqnarray}
&&D^{++}_LV^{--}_{4L}=D^{--}_LV^{++}_{4L}.\label{Vm4}
\end{eqnarray}

We define the left representation of the fourth supersymmetry generator
${\cal Q}^4_\a$ on the $\cN=3$ gauge   $U(1)$ prepotential \p{V3}
\bea
{\cal Q}^4_\a V^{++}_0=-D^0_\a V^{++}_0.\lb{4V}
\eea
In the gauge \p{VWZ}, the supersymmetry transformation contains an
additional term with the composite parameter $\La_4(\eps_4)$
\bea
\de^*(\eps_4)V^{++}_{WZ}=\eps^\a_4D^0_\a V^{++}_{WZ}-\cD^{++}\La_4(\eps_4)
\eea
and the ${\cal Q}^4_\a$-transformations of the component fields
have the form
\bea
&&\de(\eps_4)\phi^{kl}=\frac{i}2\eps^\a_4\chi^{kl}_\a,\q
\de(\eps_4)A_m=\frac{i}2(\eps_4\g_m\la),\\
&&\de(\eps_4)\la_\a=\frac12\eps^\bt_4 (\g_m)_\ab F^m,\q
\de(\eps_4)\chi^{kl}_\a=\eps^\bt_4\pa_\ab\phi^{kl}+\eps_{4\a}X^{kl},\\
&&\de(\eps_4)X^{kl}=\frac{i}2\eps^\a_4\pa_\ab\chi^{\bt kl}.
\eea

The ${\cal N}=4$ Abelian superfield strength can be expressed via the
superfield $V^{--}_{4L}$
\begin{eqnarray}
&&W^{++}_{4R}(V^{++}_L)=-\frac14(D^{++})^2V^{--}_{4L}=
[1+\te^\a_4D^0_\a-\frac14(\te_4)^2(D^0)^2]W^{++}_0(\z),\label{W4}
\end{eqnarray}
and it can be connected with the Abelian ${\cal N}=3$ analytic
superfield strength
$W^{++}_0$ \p{W3}. By definition, the superfield strength of the left
gauge superfield $W^{++}_{4R}$ satisfies the right analyticity condition
\begin{eqnarray}
&&D^{++}_\alpha W^{++}_{4R}=0,\quad D^{-+}_\alpha W^{++}_{4R}=0
\end{eqnarray}
and the harmonic condition $D^{++}_LW^{++}_{4R}=D^{++}_RW^{++}_{4R}=0$.
The superconformal transformation of this right analytic superfield has
the form
\bea
&&\de^*_4W^{++}_{4R}=2\lambda_RW^{++}_{4R}-\hat{K}_4W^{++}_{4R},\\
&&\lambda_R=\la(\z)+\frac{i}2k_m(\te^0\g^m\te_4)-U^0_{kl}b^{kl}
-\frac{i}2U^0_{kl}\eta^{(kl)}_\a+\frac{i}2\left(\te^{0\a}
-\frac12\te^\a_4\right)\eta^4_\a,
\eea
where the parameter $\la(\z)$ is defined in \p{la3}.

The superfields $V^{++}_{4L}$ and
$W^{++}_{4L}$ are defined in different ${\cal N}=4$ superspaces,
so it is convenient to use their representations $V^{++}_0$ \p{V4} and
 $W^{++}_0$ \p{W4} in
the ${\cal N}=3$ superspace. The corresponding supersymmetry transformation
of the pseudoscalar superfield strength $W^{++}_0$ can be obtained from
the ${\cal N}=4$ transformations $W^{++}_{4L}$
\bea
{\cal Q}^4_\a W^{++}_0(V^{++}_0)=D^0_\a W^{++}_0(V^{++}_0).
\eea
This transformation has a mirror form in comparison with (\ref{4V}).
This formula can be checked with the help of the relation
\bea
&&\de^*(\eps_4)V^{--}_0=-\frac12\eps^\a_4\cD^{--}D^{++}_\a V^{--}_0=
-\eps^\a_4 D^0_\a V^{--}_0
-\frac12\eps^\a_4 D^{++}_\a \cD^{--}V^{--}_0.\lb{4Vm0}
\eea

The active right $\cN=4$ superconformal transformation
of the superfield strength
\bea
&&\de^*_4W^{++}_0=2(\la-\la_4)W^{++}_0-{\cal K}_RW^{++}_0\lb{W0sc}
\eea
contains the operator ${\cal K}_R$, the $\cN=3$ parameter $\la$ \p{la3}
and the additional parameter $\la_4$ \p{lapp4}.

The analytical Abelian $\cN=3$ Chern-Simons action
\be
\int d\z^{-4} V^{++}W^{++}_0=\int d^6zdu V^{++}_0V^{--}_0\lb{CSab}
\ee
is not invariant under the $\cN=4$ supersymmetry and the
$P$-parity.

We consider the Abelian analytic pseudoscalar gauge superfield
 $A^{++}(\z)$
\bea
&&\de_\La A^{++}=-\cD^{++}\La_A,\q P A^{++}=-A^{++},\q \widetilde{A^{++}}
=-A^{++}.
\eea
The $WZ$-gauge for this superfield
\bea
&&A^{++}_{WZ}=3(\te^{++})^2U^{--}_{kl}\La^{kl}+2(\te^{++}\g^m\te^0)B_m
+2i(\te^0)^2\te^{++\a}\xi_\a+3i(\te^{++})^2\te^{0\a}U^{--}_{kl}
\rho_\a^{kl}\nn\\
&&+3i(\te^{++})^2(\te^0)^2U^{--}_{kl}Y^{kl}\lb{AWZ}
\eea
includes the scalar $\La^{kl}$, pseudoscalar $Y^{kl}$, pseudovector field
$B_m$ and spinor fields $\xi_\a, \rho^{kl}_\a$.

By definition, this $\cN=3$ superfield transforms as the right
$\cN=4$ supermultiplet
\bea
\de^*_4A^{++}=-{\cal K}_R A^{++}.
\eea
The corresponding scalar superfield strength describes the left
tensor $\cN=4$ multiplet \p{ltens}
\bea
&&L^{++}_A(A^{++})=IA^{++},
\eea
where the linear integral operator $I$ is defined in \p{W3}.

Now we can construct the $BF$ interaction with the coupling constant $\bt$
\bea
S_{BF}(V_0^{++},A^{++})=-i\bt\int d\z^{-4} V^{++}_0 L^{++}_A=
-i\bt\int d\z^{-4}du A^{++} W^{++}_0,\lb{BF}
\eea
which is invariant under the $\cN=4$ superconformal transformations and the
$P$-parity transformation.
The equivalent interaction was defined in the $\cN=4$ superspace \cite{Z7}.
The $\cN=3$ $BF$ interaction can be expressed as the difference of two Abelian
Chern-Simons terms
\bea
&&S_{BF}=-i\bt\int d\z^{-4} (V^{++}_L W^{++}_L-V^{++}_R W^{++}_R),\nn\\
&&V^{++}_0=V^{++}_L+V^{++}_R,\q A^{++}=V^{++}_L-V^{++}_R.
\eea
We note that each Chern-Simons term transforms non-trivially under the fourth
supersymmetry
\bea
&&\de^*(\eps_4)V^{++}_L=\eps^\a_4D^0_\a V^{++}_R,\q \de^*(\eps_4)V^{++}_R
=\eps^\a_4D^0_\a V^{++}_L.
\eea

The left hypermultiplet $q^+(\z)$  has the $\cN=4$
transformation
\be
\de^*_4q^+=(\la+\la_4)q^+ -{\cal K}_Lq^+.
\ee
The $\cN=4$ superconformal minimal interaction of the
left superfields  $q^+$ and $V^{++}_0$ has the form
\bea
\int d\z^{-4} \bar{q}^+\na^{++}q^+=\int d\z^{-4} \bar{q}^+(\cD^{++}q^+
+V^{++}_0q^+).
\eea

It is not difficult to construct the nonminimal $\cN=4$ superconformal
coupling of the superfield strength $W^{++}_0(V^{++}_0)$ with the right
real hypermultiplet $\Omega$
\bea
&&\int d\z^{-4}[(\cD^{++}\Omega)^2+\Omega^{-2}(W^{++}_0)^2],\\
&& \de^*_4\Omega=(\la-\la_4)\Omega
-{\cal K}_R\Omega.\lb{OmR}
\eea

The interaction of the gauge superfield $A^{++}$ with two complex
hypermultiplets
$q^{+a} (a=1,2)$ was studied  \cite{BILPSZ}
\bea
&&S(q,\bar{q},A)=\int d\z^{-4}\bar{q}^+_a(\cD^{++}+A^{++})q^{+a},\\
&&\widetilde{q^{+a}}=\bar{q}^+_a=\ve_{ab}\bar{q}^{+b},\q Pq^{+a}=
\bar{q}^{+a},\q P(\bar{q}^+_aq^{+a})=-\bar{q}^+_aq^{+a}.
\eea
The action of the Abelian ABJM model $S^0_{ABJM}=S_{BF}(V_0,A)+S(q,\bar{q},A)$
is invariant under the three nonlinear supersymmetry transformations
\cite{BILPSZ}
\bea
&&\de_\eps V^{++}_0=\frac2\bt\eps^{\a ab}\te^0_\a \bar{q}^+_aq^+_b,\q
\de_\eps A^{++}=0,\nn\\
&&\de_\eps q^{+a}=i\eps^{\a ab}[D^0_\a+\frac12D^{++}_\a A^{--}
+\te^{--}_\a L^{++}_A]q^+_b,\lb{N6}\\
&&\de_\eps \bar{q}^+_a=i\eps^\a_{ab}[D^0_\a-\frac12D^{++}_\a A^{--}
-\te^{--}_\a L^{++}_A]\bar{q}^+_b,
\nn
\eea
which form the $\cN=6$ supersymmetry together with the $\cN=3$
transformations. The algebra of these nonlinear transformations closes
on the corresponding equations of motion.

The action $S^0_{ABJM}$ is also invariant under the additional
off-shell supersymmetry \bea &&\de^*(\eps_4)V^{++}_0=\eps^\a_4D^0_\a
V^{++}_0,\q \de^*(\eps_4)A^{++}=
-\eps^\a_4D^0_\a A^{++},\nn\\
&&\de^*(\eps_4)q^{+a}=-\eps^\a_4D^0_\a q^{+a}.
\eea
These transformations do not commute with the nonlinear transformations
 \p{N6}
\bea
&&[\de^*(\eps_4),\de_\eps]V^{++}_0=-\frac1\bt\eps^\a_4\eps_\a^{ab}\bar{q}^+_a
q^+_b,\q [\de^*(\eps_4),\de_\eps]A^{++}=0,\nn\\
&&[\de^*(\eps_4),\de_\eps]q^{+a}=\eps^\a_4\eps^{\bt ab}
[-i\pa_\ab^A+\frac12D^0_\a D^{++}_\bt A^{--}
-\te^{--}_\bt D^0_\a L^{++}_A]q^+_b.
\eea

The improved $\cN=3$ left linear multiplet $w^{++}$ can be defined via
 the Abelian superfield strength
\bea
W^{++}_0(V^{++}_0)=\g(u^+_ku^+_lC^{kl} +w^{++}),\q C^{kl}C_{kl}=2,
\eea
using the constant $\g$ of dimension one and dimensionless constants
 $C_{kl}$ describing the spontaneous symmetry breaking. We introduce the
analogous improved Abelian prepotential $v^{++}$
\bea
&&V^{++}_0=3\g C^{--}(\te^{++})^2+\g v^{++},\nn\\
&&w^{++}(v^{++})=Iv^{++},\lb{shV}
\eea
where $C^{--}=u^-_ku^-_lC^{kl}$.

We define the improved coupling of  $V^{++}_0$ in the $\cN=3$ superspace
by  analogy with the  $\cN=2, d=4$ and $\cN=4, d=3$
cases \cite{GIOS,Z7}
\bea
&&S_0^L(V^{++}_0)=-\frac1\g\int d\z^{-4}\left(\frac{w^{++}}{1+
\sqrt{1+w^{++}C^{--}}}\right)^2\nn\\
&&=-\frac1\g\int dz
du\frac{v^{--}w^{++}}{(1+\sqrt{1+w^{++}C^{--}})^2}. \lb{SL} \eea
This coupling is invariant under the $\cN=4$ superconformal
transformations \bea
&&\de^*w^{++}=2(\la-\la_4)(w^{++}+C^{++})-2(\la^{++}-\la^{++}_4)C^0
-{\cal K}_Rw^{++},\lb{w4sc}
\eea
which are equivalent to transformations \p{W0sc}. The $\cN=2, d=3$
representation of the action \p{SL} was studied in \cite{BPS}.

The variation of this action in $v^{++}$ can be expressed via
the nonlinear function $B^{++}_C(w^{++})$
\bea
&&\de S_0^L(V^{++}_0)=-\frac1\g\int dz du\de v^{--}B^{++}_C(w^{++})
=-\frac1{\g}\int d\z^{-4}\de v^{++}{\cal F}^{++}_C(w^{++}),\nn\\
&&B^{++}_C(w^{++})=\frac{w^{++}}{(1+\sqrt{1+
w^{++}C^{--}})\sqrt{1+w^{++}C^{--}}},\lb{Bw}\\
&&{\cal F}^{++}_C(w^{++})=IB^{++}_C(w^{++}),\q \cD^{++}{\cal F}^{++}_C=0,
\lb{CFV}
\eea
where the operator $I$ was many times used above. The action
$S_0^L$ gives the linear equation for the function $B^{++}_C$.

The similar   $\cN=4$ superconformal coupling of the right gauge
superfield $A^{++}$ arises from the improved superfield
$l^{++}$
\bea
&&L^{++}_A(A^{++})=\g(u^+_ku^+_lc^{kl}+l^{++}), \q c^{kl}c_{kl}=2,\\
&&S_0^R(A^{++})=-\frac1\g\int d\z^{-4}\left(\frac{l^{++}}{1+
\sqrt{1+l^{++}c^{--}}} \right)^2.\lb{SR}
\eea

The combined $\cN=3$ action
\be
S^\bt(V^{++}_0,A^{++})=S_{BF}+S^L_0+S^R_0\lb{Sbt}
\ee
describes the nontrivial superconformal interaction of two Abelian
gauge superfields. The quantum properties of this model can be studied
by the method of the $\cN=3$ supergraphs \cite{BILPSZ2}.

The equivalent superconformal model was earlier considered in the
$\cN=4$ superspace \cite{Z7}. The corresponding Abelian $\cN=4$
prepotentials were defined in different mirror superspaces.
At the field-component level this model describes nonlinear
couplings of two topologically massive gauge fields with
spinor, scalar and pseudoscalar fields.

The action of the $\cN=4$ electrodynamics contains
the constant $g$ of the dimension 1/2
\begin{eqnarray}
S_2^E=-\frac1{g^2}\int d\zeta^{-4}(W^{++}_0)^2.\lb{3ED}
\end{eqnarray}
It is invariant under the  $P$-parity and $\cN=4$ supersymmetry,
but breaks the scale invariance.

We define the dimensionless uncharged analytic function of the
Abelian superfield strength, which has the right $\cN=4$ transformation
\begin{eqnarray}
&&
K=\xi^2(D^{++})^2(D^{--})^2(\cD^{--})^2(W^{++}_0)^2,\q
\de^*(\eps_4)K=-\eps^\a_4D^0_\a K,
\end{eqnarray}
where the constant $\xi$ has the dimension $-2$. The analytic
$\cN=3$ superfield density of the nonlinear electrodynamics action
\begin{eqnarray}
&&S_N^E=-\frac1{g^2}\int d\z^{-4}(W^{++}_0)^2[1+f(K)],\\
\end{eqnarray}
is proportional to the density of the quadratic action \p{3ED} and to
the nonlinear function of the superfield $K$. This action has
the $\cN=4$ supersymmetry. The component Lagrangian contains the nonlinear
terms $(F_mF^m)^n, (\pa_m\phi^{kl}\pa^m\phi_{kl})^n$.

\setcounter{equation}0
\section{Non-Abelian gauge theory}
We consider the representation $V^{++}_{4L}(\hat\z_L)=[1-\te^\a_4D^0_\a
-\frac14(\te_4)^2(D^0)^2] V^{++}(\z)$ for the non-Abelian gauge superfields.
In this case, the superconformal $\cN=4$ symmetry operator ${\cal K}_L$
\p{add} acts linearly on the non-Abelian prepotential $V^{++}$, for instance,
the fourth supersymmetry transformation has the form
${\cal Q}_\a^4V^{++}=-D^0_\a V^{++}$. The corresponding nonlinear
fourth supersymmetry transformation of the non-Abelian $\cN=3$
connection $\de^*(\eps_4)V^{--}=-\eps^\a_4\hat{Q}^4_\a V^{--}$
arises from the harmonic zero-curvature equation
 \cite{Z2}
\bea
&&\hat{Q}^4_\a V^{--}=\frac12\na^{--}D^{++}_\a V^{--}= D^0_\a V^{--}
-\frac12[D^{++}_\a V^{--},V^{--}]
+\frac12 D^{++}_\a \cD^{--}V^{--},\lb{4Vm}\\
&&\na^{++}\hat{Q}^4_\a V^{--}=\na^{--}{\cal Q}_\a^4V^{++},\quad
 \na^{\pm\pm}=\cD^{\pm\pm}
+V^{\pm\pm},\quad [\na^{++},\na^{--}]=\cD^0.
\eea
The operator $K^A_3$ \p{K3A},\p{KD} acts linearly on the superfield
$V^{--}$. The nonlinear action of the special conformal supersymmetry
generator on $V^{--}$ arises from the commutator
$\frac13(\g^m)^{\ab}[K_m,\hat{Q}^4_\bt]$.

We consider the non-Abelian $\cN=3$ superfield strength
\bea
&&\cW^{++}=-\frac14(D^{++})^2V^{--}.
\eea
The fourth supersymmetry transformation of this superfield has the form
\bea
&&\hat{Q}^4_\a \cW^{++}=-\frac14(D^{++})^2\hat{Q}^4_\a V^{--}=
D^0_\a\cW^{++}-[D^{++}_\a V^{--},\cW^{++}],\lb{4QW}
\eea
it preserves analyticity
\bea
&&D^{++}_\bt\hat{Q}^4_\a \cW^{++}=0,\q D^{++}_\bt D^{++}_\a V^{--}=
-2\ve_{\bt\a}\cW^{++}.
\eea
We note that the nonlinear terms in the transformations \p{4Vm} and \p{4QW}
differ by the coefficient 2.

The action of the $\cN=3$ Yang-Mills theory has the form
\bea
&&S_{SYM}=\frac1{4g^2}\int d\z^{-4}(D^{++})^2\mbox{Tr}\,V^{--}
\cW^{++}=-\frac1{g^2}\int d\z^{-4}\mbox{Tr}\,(\cW^{++})^2.
\eea
The gauge-invariant analytic density of the action
$L^{(+4)}=\mbox{Tr}\,(\cW^{++})^2$ transforms linearly by
analogy with the Abelian quantity $(W^{++}_0)^2$ \p{W0sc}
\bea
&&\de^*_4L^{(+4)}=4(\la-\la_4)L^{(+4)}-{\cal K}_RL^{(+4)},
\eea
so the action $S_{SYM}$ is invariant under the fourth supersymmetry
 $Q^4_\a L^{(+4)}=D^0_\a L^{(+4)}$, although it breaks the conformal
invariance.

Using the improved Abelian right tensor multiplet $w^{++}(V^{++}_0)$
\p{w4sc} we can construct the right analytic density
 $F(w)=(1+w^{++}C^{--})^{-3/2},$
\bea
&& \de^*_4F(w)=-2(\la-\la_4)F(w) +\cD^{++}A^{--}-{\cal
K}_RF(w), \eea
where $A^{--}$ is some analytic superfield, which is
series in degrees of $w^{++}C^{--}$ and is linear in superconformal
parameters $\la-\la_4$ and $ \la^{++}-\la^{++}_4$. This density
allows us to define the superconformal generalization of the
non-Abelian gauge action
\bea
&&S(V^{++},V^{++}_0)=-\frac1{g^2}\int
d\z^{-4}F(w) \mbox{Tr}\,(\cW^{++})^2,\lb{NA}
\eea
which contains also the Abelian gauge superfield $w^{++}$. To prove the
superconformal symmetry we use transformations of $F(w)$ and
$L^{(+4)}$, and formulas \bea &&-\int d\z^{-4}{\cal
K}_RL^{(+4)}=-2\int d\z^{-4}(\la-\la_4)L^{(+4)}, \q
\cD^{++}L^{(+4)}=0.
\eea

The superconformal generalization of the non-Abelian theory can also be
constructed with the help of the right hypermultiplet
 $\Omega(\z)=\gamma C^{kl}U^0_{kl}
+\omega(\z)$ \p{OmR}, then the superconformal density is
$\Omega^{-2}\mbox{Tr}\,(\cW^{++})^2$. The kinetic term for the superfield
$\Omega$ has a standard form. The parameters $\gamma$ and $C^{kl}$
describe the spontaneous breaking of the superconformal symmetry
in this interaction.

\setcounter{equation}0
\section{Conclusions}
We review the superconformal transformations in the standard
and analytic harmonic superspaces with the $\cN=3, d=3$ supersymmetry.
The active local form of the $\cN=3$  transformation
is defined via the Killing operator $K_3$ which contains the superconformal
generators. The commutators
of $K_3$ with the flat $\cN=3$ spinor derivatives determine matrices
of the superconformal transformations. These matrices are used in
the  superconformal transformations of the standard $\cN=3$ superfields.
The $\cN=3$ analytic superspace is convenient for the description
of the hypermultiplet couplings with the gauge Chern-Simons or
Yang-Mills superfields. Analogous superconformal structures
were considered in the $\cN=4, d=3$ superspaces \cite{Z7},
but this formalism has difficulties in the analysis of the left-right $\cN=4$
supermultiplet couplings.

We study the $\cN=4$ superconformal models in the framework of
more flexible  $\cN=3$ harmonic superspace. Left and right superfields
from the mirror $\cN=4$ superspaces are connected by the operator transformations
with the corresponding $\cN=3$ harmonic superfields in the same superspace.
The representations of the additional  $\cN=4$ superconformal generators
$Q^4_\a, S^4_\a$ and $A_{kl}$ are constructed in the $\cN=3$
analytic superspace. The mirror map changes signs of the corresponding
$\cN=4$ superconformal transformations in the $\cN=3$ superspace.

It is easy to reformulate the $\cN=4$ superfield models \cite{Z7}
in the $\cN=3$ superfield representation analyzing the additional
supersymmetry. We prove that the gauge prepotential $V^{++}_0$ and
its superfield strength are the mirror $\cN=4$ supermultiplets.
The superfield Abelian $\cN=3$ $BF$ coupling $S_{BF}$ \p{BF} connects
the left scalar gauge prepotential with the superfield strength of the
right pseudoscalar gauge prepotential. This $BF$ coupling is  part of
the $U(1)\times U(1)$ ABJM model in the $\cN=3$ superspace  which has
the on-shell $\cN=6$ supersymmetry \cite{BILPSZ}. The improved
superconformal forms of the left and right tensor multiplets $S_0^L$
\p{SL} and $S^R_0$ \p{SR} are defined in the same $\cN=3$ superspace.
We derive the $\cN=3$ representation of the superfield equations of motion
for the interesting model based on the action $S_{BF}+S^L_0+S^R_0$.

We consider the $\cN=4$ supersymmetry transformations on the non-Abelian
$\cN=3$ superfields and construct the superconformal coupling of these
superfields with the Abelian gauge superfield. A similar coupling
was considered earlier in the $\cN=4$ superspace \cite{Z7}.

\renewcommand\theequation{A.\arabic{equation}}
\setcounter{equation}0
\section*{Appendix}

The  $ \cN=4, d=3$ superspace is covariant with respect to the
Lorentz group $SO(2,1)\sim SL(2,R)$ and the automorphism group
$SU_L(2)\times SU_R(2)$. The important property of the $\cN=4$
superspace is the discrete  symmetry with respect to the mirror map
\be \cM:\hspace{1cm}SU_L(2)~\leftrightarrow~SU_R(2)\lb{Msym} \ee We
consider the coordinates of the $d=3, \cN=4$ superspace in the
central basis \cite{Z3,Z5,Z7}: \bea &&z=(x^m, \te^\a_{ka}), \eea
where $i$ and $a$ are the two-component  indices of the automorphism
groups $SU_L(2)$ and $SU_R(2)$, respectively.

In this paper, we identify indices of two $SU(2)$ groups
and consider the following decomposition of the $\cN=4$ spinor
coordinates:
\bea
&&\te^\a_{ka}~\rightarrow ~\te^\a_{kl}=\te^\a_{(kl)}+\frac12
\ve_{kl}\te^\a_4,
\eea
where $\te^\a_{(kl)}$ are the $\cN=3$ $CB$ coordinates and $\te^\a_4$
is an additional fourth spinor coordinate.
The $\cN=3$ central basis in the
$\cN=4$ superspace has the form
\bea
&&\hat{z}=(x^m, \te^\a_{(kl)}, \te^\a_4).
\eea

We define the corresponding decomposition of the partial spinor derivatives
\bea
&&\pa_\a^{kl}=\pa_\a^{(kl)}-\ve^{kl}\pa^4_\a,\q
\pa_\a^{kl}\te^\bt_{jn}=\de^\bt_\a\de^k_j\de^l_n ,\nn\\
&&\pa_\a^{(kl)}\te^\bt_{(jn)}=\frac12\de^\bt_\a(\de^k_j\de^l_n+\de^k_n\de^l_j),
\q \pa^4_\a\te^\bt_4=\de^\bt_\a
\eea
and the $\cN=4$ spinor derivatives
\bea
&&D^{kl}_\a=D^{(kl)}_\a-\ve^{kl}D^4_\a,\q D^4_\a=
\pa^4_\a+\frac{i}2\te^\g_4\pa_{\a\g}.
\eea

The $\cN=3$ superspace is invariant under the mirror map
\bea
&&\cM\te^\a_{kl}=\te^\a_{lk},\q\cM \te^\a_{(kl)}=\te^\a_{(kl)},\q
\cM \te^\a_4=-\te^\a_4.\lb{MCB}
\eea

The
$\cN=4$ Killing operator contains the corresponding superconformal
parameters and generators
\bea
&&K_4=c^mP_m+l^m\hat{M}_m+b\hat{D}+k^m\hat{K}_m+\om^{kl}L_{kl}+\Om^{kl}R_{kl}
\nn\\
&&
+\eps^\a_{(kl)}Q^{(kl)}_\a+\eps^\a_4Q^4_\a+\eta^\a_{(kl)}\hat{S}^{(kl)}_\a
+\eta^\a_4S^4_\a.
\eea
Generators of the  $SU_L(2)\times SU_R(2)$ transformations
$L_{kl}=\frac12(V_{kl}+A_{kl})$ and
$R_{kl}=\frac12(V_{kl}-A_{kl})$ can be written in terms of the
$SU_V(2)$ generator $V_{kl}$ \p{3V} and the additional generator
$A_{kl}=\te^\a_4\pa_{\a(kl)}-2\te^\a_{(kl)}\pa^4_\a$ which connects spinor
coordinates $ \te^\a_{(kl)}$ and $\te^\a_4$. Now we can separate the
operator of the $\cN=3$ superconformal transformations $K_3$ and the
operator of additional transformations $K_{4/3}$
\bea
&&K_4=K_3+K_{4/3},\q K_{4/3}=b^{kl}A_{kl}+\eps^\a_4Q^4_\a+\eta^\a_4S^4_\a.
\eea

The $\cN=4$ superconformal transformations of the spinor coordinates
have the form
\bea
&&\de\te^\a_{kl}=\eps^\a_{kl}+l^\a_\g\te^\g_{kl}+\frac12b\,\te^\a_{kl}
-\om_k^j\te^\a_{jl}-\Om_l^j\te^\a_{kj}-\frac12k^\bt_{\g}x^\a_\bt\te^\g_{kl}
-\frac{i}4[\T+\frac12(\te_4)^2]\te^\g_{kl} k^\a_\g\nn\\
&&-\frac12x^\a_\g\eta^\g_{kl}+i\eta^{jn\g}\te^\a_{jn}\te_{kl\g}-\frac{i}4
[\T+\frac12(\te_4)^2]\eta^\a_{kl}=K_4\te^\a_{kl}
\eea
where $\T=\te^\a_{(kl)}\te^{(kl)}_\a,\q\T+\frac12(\te_4)^2=
\te^\a_{kl}\te^{kl}_\a$.

The fourth supersymmetry generator contains $\te^\g_4$ and the corresponding
spinor derivative
\bea
&&Q^4_\a=\pa^4_\a-\frac{i}2\te^\g_4\pa_{\a\g}.
\eea

The mirror map \p{MCB} yields the automorphism of the $\cN=4$ superconformal
Lie superalgebra
\bea
\cM A_{kl}=-A_{kl},\q \cM Q^4_\a=-Q^4_\a,\q \cM S^4_\a=-S^4_\a,\q
\cM K_3=K_3.
\eea

The left analytic $\cN=4$ basis \cite{Z7} uses the left $SU(2)_L/U(1)$
harmonics $u^\pm_k$ and the coordinates
\bea
&&\z_L=(x^m_L, \te^{+\a}_l, u^\pm_i),\q \te^{-\a}_l,\q \te^{\pm \a}_l=
-u^{\pm k}\te^\a_{kl},\q \pa^{\mp k}_\bt\te^{\pm \a}_l=\de^k_l\de^\a_\bt,
\nn\\
&&x^m_L=x^m+iu^{+j}u^{-k}\ve^{ln}(\g^m)_\ab\te^{\a}_{jn}
\te^{\bt}_{kl},\q \pa^L_nx^m_L=\delta^m_n,\lb{xLAB}\\
&&u^+_lD^{kl}_\a=D^{+k}_\a=\pa^{+k}_\a,\q
u^-_lD^{kl}_\a=D^{-k}_\a=-\pa^{-k}_\a+2i\te^{-k\bt}\pa^L_\ab.
\eea

We consider the alternative representation of the left analytic basis
\bea
&&\hat\z_L=(x^m_L, \te^{++\a},\te^{+-\a},u^\pm_i),\q
\te^{\pm\pm\a}=u^{\pm k}u^{\pm l}\te^\a_{kl},\q
\te^{\pm\mp\a}=u^{\pm k}u^{\mp l}\te^\a_{kl},\lb{LAB}\\
&&x^m=x^m_L-i(\g^m)_\ab\te^{++\a}\te^{--\bt}+i(\g^m)_\ab
\te^{+-\a}\te^{-+\bt},\\
&&\te^{+\a}_l=
u^+_l\te^{+-\a}-u^-_l\te^{++\a},\q
\te^{-\a}_l=u^+_l\te^{--\a}-u^-_l\te^{-+\a}.
\eea
It is easy to connect the partial derivatives in different representations,
for instance,
\bea
&&\pa^{+l}_\a=-u^{-l}\pa^{++}_\a-u^{+l}\pa^{+-}_\a,\q
\pa^{++}_L=\pa^{++}+\te^{-+\a}\pa^{++}_\a
+\te^{++\a}\pa^{-+}_\a.
\eea

The alternative representations of the $\cN=4$ spinor and harmonic derivatives
have the form
\bea
&&D^{++}_\a=\pa^{++}_\a,\q D^{+-}_\a=-\pa^{+-}_\a,\nn\\
&&D^{-+}_\a=-\pa^{-+}_\a+2i\te^{-+\bt}\pa^L_\ab,\q
D^{--}_\a=\pa^{--}_\a+2i\te^{--\bt}\pa^L_\ab,\\
&&D^{++}_L=\pa^{++}+2i\te^{++\a}\te^{+-\bt}\pa^L_\ab
+\te^{++\a}(\pa^{+-}_\a+\pa^{-+}_\a)+(\te^{+-\a}+\te^{-+\a})\pa^{++}_\a,\nn\\
&&D^{--}_L=\pa^{--}-2i\te^{--\a}\te^{-+\bt}\pa^L_\ab
+\te^{--\a}(\pa^{-+}_\a+\pa^{-+}_\a)+(\te^{-+\a}+\te^{-+\a})\pa^{--}_\a,\\
&&D^0_L=\pa^0+2\te^{++\a}\pa^{--}_\a
-2\te^{--\a}\pa^{++}_\a.\nn
\eea
where we use the partial derivatives in the new coordinates
\bea
&&
\pa^{\mp\mp}_\a\te^{\pm\pm\bt}=\pa^{\mp\pm}_\a\te^{\pm\mp\bt}=\de^\bt_\a,\q
\pa^{\pm\pm}\te^{\pm\pm\bt}=0.
\eea

In the left basis, the  $\cN=4$ supersymmetry generators $Q^{kl}_\a=Q^{(kl)}_\a
-\ve^{kl}Q^4_\a$ have the form
\bea
&&Q^{(kl)}_\a=U^{++kl}\pa^{--}_\a+U^{--kl}(\pa^{++}_\a
-2i\te^{++\bt}\pa^L_\ab)+U^{0kl}(\pa^{-+}_\a
+\pa^{+-}_\a+2i\te^{+-\bt}\pa^L_\ab),\nn\\
&&Q^4_\a=\frac12\left(\pa^{-+}_\a-\pa^{+-}_\a\right)-i\te^{+-\bt}\pa^L_\ab.
\lb{QA4}
\eea
It is also easy to construct the generator of the special conformal
transformations
$K_m={\cal K}^L_m+K^\prime_m$
\bea
&&{\cal K}^L_m=-i(\te^{++}\g_m\te^{+-})\pa^{--}+x_{mL}x^n_L\pa^L_n
-\frac12(x_L)^2\pa^L_m
-\frac{i}2(\g_m)^\a_\g\te^{++\g}(\te^{+-})^2\pa^{--}_\a\nn\\
&&+\frac12x^n_L(\g_n)^\ab(\g_m)_{\bt\g}[\te^{++\g}\pa^{--}_\a
+\te^{+-\g}\pa^{-+}_\a],\\
&&
K^\prime_m=\frac12x^n_L(\g_n)^\ab(\g_m)_{\bt\g}[\te^{-+\g}\pa^{+-}_\a
+\te^{--\g}\pa^{++}_\a]
-\frac{i}2(\g_m)^\a_\g\te^{+-\g}(\te^{-+})^2\pa^{+-}_\a
\nn\\&&
+i(\g_m)_{\bt\g}\te^{++\bt}(\te^{-+\g}-\te^{+-\g})\te^{--\a}\pa^{+-}_\a
+\frac{i}2(\g_m)^\a_\g\te^{++\g}(\te^{--})^2\pa^{++}_\a,\nn
\eea
where the operator ${\cal K}^L_m$ acts on the left analytic superfields.

We consider representations of the $SU_L(2)\times SU_R(2)$ superconformal generators
in this basis
\bea
&&L_{kl}=U^{++}_{kl}\pa^{--}+U^{++}_{kl}\te^{+-\a}\pa^{--}_\a
+U^{++}_{kl}\te^{--\a}\pa^{+-}_\a\nn\\
&&+U^0_{kl}(\te^{++\a}\pa^{--}_\a+\te^{+-\a}\pa^{-+}_\a)
-U^0_{kl}(\te^{--\a}\pa^{++}_\a+\te^{-+\a}\pa^{+-}_\a)\nn\\
&&+U^{--}_{kl}(\te^{+-\a}\pa^{++}_\a+\te^{++\a}\pa^{+-}_\a)
+2iU^{--}_{kl}\te^{++\a}\te^{+-\a}\pa^L_\ab,\lb{4LL}\\
&&R^{kl}=-U^{++kl}\te^{+-\a}\pa^{--}_\a+U^{0kl}\te^{++\a}\pa^{--}_\a
-U^{0kl}\te^{+-\a}\pa^{-+}_\a+U^{--kl}\te^{++\a}\pa^{-+}_\a\nn\\
&&-U^{++kl}\te^{--\a}\pa^{+-}_\a-U^{0kl}\te^{--\a}\pa^{++}_\a
+U^{0kl}\te^{-+\a}\pa^{+-}_\a+U^{--kl}\te^{-+\a}\pa^{++}_\a.\lb{4LR}
\eea
We can study the active $\cN=4$ superconformal transformations
of the left analytic superfields $\Phi_L(\hat\z_L)$, for instance,
\bea
&&\de^*\Phi_L=-[\om^{kl}L_{kl}+\Om^{kl}R_{kl}+\eps^\a_{kl}Q^{kl}_\a]
\Phi_L\nn\\
&&
=-[a^{kl}V_{kl}+b^{kl}A_{kl}+\eps^\a_{(kl)}Q^{(kl)}_\a+\eps^\a_4Q^4_\a]
\Phi_L.
\eea

The superconformal operators $R_{kl}$ do not act on the left even coordinates
\bea
&&R_{kl}x^m_L=0,\q R_{kl}u^\pm_i=0,\q [ [R_{kl},D^{\pm\pm}_L]=0.
\eea

The $\cN=4$ Killing operator $K_4$ satisfies the following relations
in the left basis:
\bea
&&[K_4,D^{++}_L]=-\la^{++}_LD^0_L,\q [K_4,D^{--}_L]=-(D^{--}_L\la^{++}_L)
D^{--}_L,\\
&&\la^{++}_L=(a^{kl}+b^{kl})u^+_ku^+_l-i(\te^{++}\g^m\te^{+-})k_m
\nn\\
&&+i(\te^{+-\a}u^+_ku^+_l-\te^{++\a}u^-_ku^+_l)\eta^{(lk)}_\a
-\frac{i}2\te^{++\a}\eta^4_\a.\lb{laL}
\eea

We obtain the coordinates and partial derivatives of the right analytic
$\cN=4$  basis $\hat\z_R=\cM\hat\z_L$  using the mirror map from the left basis
 \p{LAB}
\bea
&&\hat\z_R=\cM\hat\z_L=(x^m_R, \te^{++\a},\te^{-+\a}, u^\pm_i),
\lb{RAB}\\
&&x^m_R=\cM x^m_L=x^m+iu^{+j}u^{-k}\ve^{ln}(\g^m)_\ab\te^{\a}_{nj}
\te^{\bt}_{lk}\nn\\
&&
=x^m_L+2i(\g^m)_\ab\te^{+-\a}\te^{-+\bt},\lb{xRAB}\\
&&\cM\te^{\pm\pm\a}=\te^{\pm\pm\a},\q \cM\te^{\pm\mp\a}=\te^{\mp\pm\a},
\q \cM u^\pm_k=u^\pm_k,\\
&&\cM \pa^L_m=\pa^R_m,\q \cM\pa^{\pm\pm}_\a=\hat\pa^{\pm\pm}_\a,\q
\cM\pa^{\pm\mp}_\a=\hat\pa^{\mp\pm}_\a,\q \cM\pa^{\pm\pm}
=\pa^{\pm\pm}.\nn
\eea
The right basis in  \cite{Z7} contains the independent right
harmonics  $v^{(\pm)}_a$. We consider the active transformations
of superfields and, therefore, can formally use the same harmonics
and partial harmonic derivatives in the left and right bases.
In the right basis, the spinor and harmonic derivatives have the
form
\bea
&&\hat{D}^{++}_\a=\hat\pa^{++}_\a=\cM D^{++}_\a,\q
\hat{D}^{-+}_\a=-\hat\pa^{-+}_\a=\cM D^{+-}_\a,\\
&&\hat{D}^{+-}_\a=-\hat\pa^{+-}_\a+2i\te^{+-\bt}\pa^R_\ab=\cM D^{-+}_\a,\q
\hat{D}^{--}_\a=\hat\pa^{--}_\a+2i\te^{--\bt}\pa^R_\ab=\cM D^{--}_\a,\\
&&D^{++}_R=\pa^{++}+2i\te^{++\a}\te^{-+\bt}\pa^R_\ab
+\te^{++\a}(\hat\pa^{-+}_\a+\hat\pa^{+-}_\a)+(\te^{+-\a}+\te^{-+\a})
\hat\pa^{++}_\a=\cM D^{++}_L,\nn\\
&&D^{--}_R=\pa^{--}-2i\te^{--\a}\te^{+-\bt}\pa^R_\ab
+\te^{--\a}(\hat\pa^{-+}_\a+\hat\pa^{+-}_\a)+(\te^{+-\a}+\te^{-+\a})
\hat\pa^{--}_\a.
\eea

The right analytic part of the generator for the special conformal
transformations has the form
\bea
&&{\cal K}^R_m=\cM{\cal K}^L_m=-i(\te^{++}\g_m\te^{-+})\pa^{--}
+x_{mR}x^n_R\pa^R_n-\frac12(x_R)^2\pa^R_m
-\frac{i}2(\g_m)^\a_\g\te^{++\g}(\te^{-+})^2\hat\pa^{--}_\a\nn\\
&&+\frac12x^n_R(\g_n)^\ab(\g_m)_{\bt\g}[\te^{++\g}\hat\pa^{--}_\a
+\te^{-+\g}\hat\pa^{+-}_\a].
\eea

The right representation of the $SU_L(2)\times SU_R(2)$ generators and the
supersymmetry generators can be obtained by the mirror map of the left
representation
\bea
&&\hat{L}_{kl}=\cM R_{kl}=-U^{++kl}\te^{-+\a}\hat\pa^{--}_\a
+U^{0kl}\te^{++\a}\hat\pa^{--}_\a
-U^{0kl}\te^{-+\a}\hat\pa^{+-}_\a+U^{--kl}\te^{++\a}\hat\pa^{+-}_\a\nn\\
&&-U^{++kl}\te^{--\a}\hat\pa^{-+}_\a-U^{0kl}\te^{--\a}\hat\pa^{++}_\a
+U^{0kl}\te^{+-\a}\hat\pa^{+-}_\a+U^{--kl}\te^{+-\a}\hat\pa^{++}_\a,\nn\\
&& \hat{R}_{kl}=\cM L_{kl},\q
\hat{A}_{kl}=-\cM A_{kl}=\cM(R_{kl}-L_{kl}),\\
&&\hat{Q}^{(kl)}_\a=U^{++kl}\hat\pa^{--}_\a+U^{--kl}(\hat\pa^{++}_\a
-2i\te^{++\bt}\pa^R_\ab)+U^{0kl}(\hat\pa^{-+}_\a
+\hat\pa^{+-}_\a+2i\te^{-+\bt}\pa^R_\ab),\nn\\
&& \hat{Q}^4_\a=-\cM Q^4_\a=\frac12\left(\hat\pa^{-+}_\a
-\hat\pa^{+-}_\a\right)+i\te^{-+\g}\pa^R_{\a\g}.
\eea
The right representation of the Killing operator
 $\hat{K}_4=K_3-K_{4/3}$ satisfies the relation
\bea
&&[\hat{K}_4,D^{++}_R]=-\la^{++}_RD^0_R,\\
&&\la^{++}_R=\cM\la^{++}_L=(a^{kl}-b^{kl})u^+_ku^+_l-i(\te^{++}\g^m
\te^{-+})k_m\nn\\
&&+i(\te^{-+\a}u^+_ku^+_l-\te^{++\a}u^-_ku^+_l)\eta^{(lk)}_\a
+\frac{i}2\te^{++\a}\eta^4_\a.\lb{laR}
\eea
\\

{\bf  Acknowledgements.} The author is grateful to E.A. Ivanov for
the interesting discussions. The work is partially supported by
RFBR grants N  09-02-01209,  09-01-93107-CNRS and 09-02-91349-DFG,
by  grant DFG 436 RUS 113/669/0-4R and by the Heisenberg-Landau
programme.

\end{document}